\documentclass[10pt,journal,cspaper,compsoc]{IEEEtran}
\ifCLASSOPTIONcompsoc
\else
\fi
\ifCLASSINFOpdf
\else
\fi
\usepackage[boxed,algosection,linesnumbered]{algorithm2e}
\usepackage{epsfig}
\usepackage{bbm}
\usepackage{fullpage}
\usepackage{amssymb}
\usepackage{amsmath}
\usepackage{amscd}
\usepackage{amsthm}
\begin{document}
%
\title{On Dynamic Job Ordering and Slot Configurations  for Minimizing the Makespan Of Multiple MapReduce Jobs}
%
%
%
%
\author{Wenhong~Tian,~\IEEEmembership{}
Guangchun~Luo, Ling~Tian, ~\IEEEmembership{}
and~Aiguo~Chen ~\IEEEmembership{}
\IEEEcompsocitemizethanks{\IEEEcompsocthanksitem Prof. Tian is in the School of Information and Software Engineering, University of Electronic Science and Technology of China. E-mail: tian\_wenhong@uestc.edu.cn;This research is sponsored by the National Science Foundation of China with project ID 61450110440;
\IEEEcompsocthanksitem G.Luo and L. Tian and A. Chen are with the same University.}
\thanks{}}
%
%
\markboth{IEEE Transactions on Service Computing
,~Vol.~, No.~, ~2016}%
{Shell \MakeLowercase{\textit{et al.}}: Bare Demo of IEEEtran.cls for Computer Society Journals}
%
\IEEEcompsoctitleabstractindextext{%
\begin{abstract}
MapReduce is a popular parallel computing paradigm for Big Data processing in clusters and data centers. It is observed that different job execution orders and MapReduce slot configurations for a MapReduce workload have significantly different performance with regarding to the makespan, total completion time, system utilization and other performance metrics. There are quite a few algorithms on minimizing makespan of multiple MapReduce jobs. However, these algorithms are heuristic or suboptimal. The best known algorithm for minimizing the makespan is 3-approximation by applying Johnson rule. In this paper, we propose an approach called UAAS algorithm to meet the conditions of classical Johnson model. Then we can still use Johnson model for an optimal solution. We explain how to adapt to Johnson model and provide a few key features of our proposed method.
\end{abstract}
}
\maketitle
\IEEEdisplaynotcompsoctitleabstractindextext
%
\IEEEpeerreviewmaketitle
\section{Introduction}
With the rapid increase in size and number of jobs that are being processed in the MapReduce framework, efficiently scheduling multiple jobs under this framework is becoming increasingly important. Job scheduling in MapReduce framework brings a new challenge to Cloud computing [1] such as minimizing the makespan, load balancing and reduce data skew etc., it has already received much attention.
Originally, MapReduce was designed for periodically running large batch workloads with a FIFO (First-In-First-Out) scheduler. As the number of users sharing the same MapReduce cluster increased, there are Capacity scheduler [11] and Hadoop Fair Scheduler (HFS) [7] which intend to support more efficient cluster sharing. There are also a few research prototypes of Hadoop schedulers that aim to optimize explicitly some given scheduling metrics, e.g., FLEX [8], ARIA [4]. A MapReduce simulator called SimMR [5] is also developed to simulate different workload and performance of MapReduce.~Yao et al. [15] proposed a scheme which uses slot ratio between Map and Reduce tasks as a tunable knob for dynamically allocating slots. However, as pointed out in [1], the existing schedulers do not provide a support for minimizing the makespan for a set of jobs.\\
Starfish project [9] proposes a workflow-aware scheduler that correlate data (block) placement with task scheduling to optimize the workflow completion time. Zhao et al. [16] propose a reference service framework for integrating scientific workflow management systems into various cloud platforms.
Moseley et al. [10] formulate MapReduce scheduling as a generalized version of the classical two-stage flexible flow-shop problem with identical machines; they provide a 12-approximation algorithm for the offline problem of minimizing the total flow-time, which is the sum of the time between the arrival and the completion of each job. Zhu et al. [15] consider nonpreemptive case to propose $\frac{3}{2}$-approximation for offline scheduling regarding the makespan where they did not considering job ordering or applying Johnson model. In [1] and [2], the authors propose heuristics to minimize the makespan, the proposed algorithm called BalancedPools by considering two pools for a Hadoop cluster. Tang et al. [17] proposed a new algorithm called MK$\_$JR for minimizing the makespan. The works of  [1] and [17] are closely related to our research in minimizing the makespan. However, our present work meets all the requirements of Johnson model and provide optimal solution to offline scheduling while Verma et al. [1] did not modify Johnson's model and provided separating pools (called BalancedPools) for minimizing the makespan, and BalancedPools is a heuristic approach but not optimal in many cases. MK$\_$JR is a 3-approximation algorithm for minimizing the makespan. There is still room for improving the performance of MapReduce regarding minimize the makespan. \\

In summary, there is only a small number of scheduling algorithms with regarding to minimize the makespan of a set of MapReduce jobs in open literature and still much room for improving the performance of MapReduce regarding minimizing the makespan.
Therefore, we propose new modeling and scheduling approaches for offline jobs in the following sections.
The major contributions of this paper include: \\
1) provided a new modeling and scheduling approach for multiple MapReduce jobs; \\
2) proposed an optimal algorithm for offline scheduling considering Map and Reduce stages by adapting to classical Johnson's model; \\
3) introduced a few key features (theorems) of our proposed algorithm (UAAS).

\section{Problem Formulation}
We consider the following problem as in [1] [17]. Let $J$=
$\{J_1 , J_2 ,\ldots, J_n \}$ be a set of $n$ MapReduce jobs with no data dependencies between them. These jobs can be executed in any order. A MapReduce job $J_i$ consists of two stages, a map stage $M$ and reduce stage $R$. Each stage consists of a number of tasks. The workload is executed on a MapReduce cluster under FIFO scheduling by default, consisting of a set of (map and reduce) slots. Let $S^M$ and $S^R$ denote the set of map slots and reduce slots configured by MapReduce administrator (i.e., $S$=$S^M ~U~ S^R$), so that the number of map slots and reduce slots are $|S^M|$ and $|S^R|$, correspondingly.
Let $\phi$ denote the job submission order for a MapReduce workload.
We consider the offline case in which all the jobs are available at time 0. Let $c_i$ denote the completion time of $J_i$ (i.e., the time when $J_i$'s reduce tasks all finish). The makespan for the workload $\{J_1 , J_2 ,\ldots, J_n \}$ is defined as $C_{max}$ =$max_{i\in [n]} c_i$. 

We denote $|J_i^M|$ and $|J_i^R|$ as the number of tasks in $J_i$'s map stage and reduce stage, respectively. Let $t_{i,j}^M$ and $t_{i,j}^R$ denote the execution time of $J_i$'s $j$th map task and jth reduce task, respectively. Let $T_i^M$ and $T_i^R$ denote the execution time of $J_i$'s map and reduce stage respectively. 
$J_i$ requests $S_i^M\times S_i^R$ MapReduce slots and has Map and Reduce stage durations ($T_i^M, T_i^R$) respectively. The system scheduler can change a job's MapReduce slots allocation depending on available resources. We aim to determine an
order (a schedule) of execution of jobs $J_i \in J$ such that the
makespan of all jobs is minimized. Let us set the actually allocated MapReduce slots for job $J_i$ as $|A_i^M| \times
|A_i^R|$, the max available MapReduce slots in the Hadoop cluster is $|S_i^M| \times |S_i^R|$. The original Johnson Rule [3] considers that “There are $n$ items which must go through one production stage or machine and then a second one. There is only one machine for each stage. At most one item can be on a machine at a given time”. We consider MapReduce as two non-overlapped stages, i.e., map and reduce stage respectively, the same as in [1][17]. Also we classify all jobs into Map type and Reduce type. For Map type jobs, their map durations should be smaller than reduce durations while Reduce type jobs have longer reduce durations than map durations. Based on these assumptions and Johnson algorithm [1], we can obtain the optimal makespan of a set of jobs as follows:
\begin{equation}
C_{max}=\sum_{i=1}^{n} T_i^R+max_{u=1}^{n} K_u
\end{equation}
where
\begin{equation}
K_u=\sum_{i=1}^{u} T_i^M-\sum_{i=1}^{u-1} T_i^R.
\end{equation}
\textbf {Observation 1. If each job utilizes either all map or all reduce slots during its processing, there is a perfect match between the assumptions of the classic Johnson algorithm for two-stage production system and MapReduce job processing, then Johnson's algorithm can be applied to find optimal solution for minimizing the makespan of a set of MapReduce jobs.} \\
Based on our observations and intensive real test experiences, we propose a new method called UAAS (Utilizing All Available Slots) algorithm, with the pesudocode given in Algorithm 2.1.
The following theorem is the key strategy for our results. \\
\textbf {Theorem 1. Based on available MapReduce slots in the system, the scheduler can increase or decrease the number of MapReduce slots to the job to meet the requirements of JohnSon Rule, the result obtained by UAAS algorithm following Johnson rule is optimal regarding to minimize the makespan. } \\

\textbf {Proof}: The original Johnson Rule [3] considers that "there are $n$ items which must go through one production stage or machine and then a second one. There is only one machine for each stage. At most one item can be on a machine at a given time". To adapt the MapReduce model, we treat the Map and Reduce stage resources as a whole (like a single machine), i.e., to represent the resources as MapReduce slots in the whole in our algorithm UAAS. USSA algorithm allocates all available MapReduce slots to each job at each stage, so that UAAS meets all requirements of Johnson Rule. Since Johnson Rule obtains optimal results with regarding to minimize the makespan (the proof is provided in [3]), and our UAAS algorithm meets all requirements of Johnson Rule, therefore UAAS obtains the optimal result with regard to minimizing the makespan. $\blacksquare$ \\
\begin{algorithm}
\caption{Utilizing All Available Slots (UAAS) algorithm}
\label{algorithm_mffde1}
\SetKwInOut{Input}{input}
\SetKwInOut{Output}{output}
\Input{the total number of MapReduce slots $(|S^M|, |S^R|)$ for a Hadoop cluster, estimated all Jobs' Map and Reduce durations $(T_i^M,T_i^R)$ [1] by utilizing all available Map and Reduce slots for each job in the cluster}
\Output{the scheduled jobs, the makespan $C_{max}$}
List the Map and Reduce's durations in two vertical columns (implemented in a list) \;
\For{ all $J_i\in J$}{
Find the shortest one among all durations (min ($T_i^M$, $T_i^R$))\;
~In case of ties, for the sake of simplicity, order the item with the smallest subscript first. In case of a tie between Map and Reduce, order the item according to the Map \;
IF it is the first job of Map type, place the corresponding item at the first place \;
ELSE it is the first job of Reduce type, place the corresponding item at the last place \;
IF it is Map type job (and not the first job), place the corresponding item right next to the previous job (i.e., in non-decreasing order of Map durations) \;
ELSE it is Reduce type job (and not the first job), place the corresponding item left next to the previous job (i.e., in non-increasing order of Reduce durations) \;
Remove both durations for that job \;
Repeat these steps on the remaining set of jobs \ 
}
Compute the makespan ($C_{max}$)\
\end{algorithm}

\begin{figure} [htp!]
\begin{center}
{\includegraphics [width=0.42\textwidth,angle=-0] {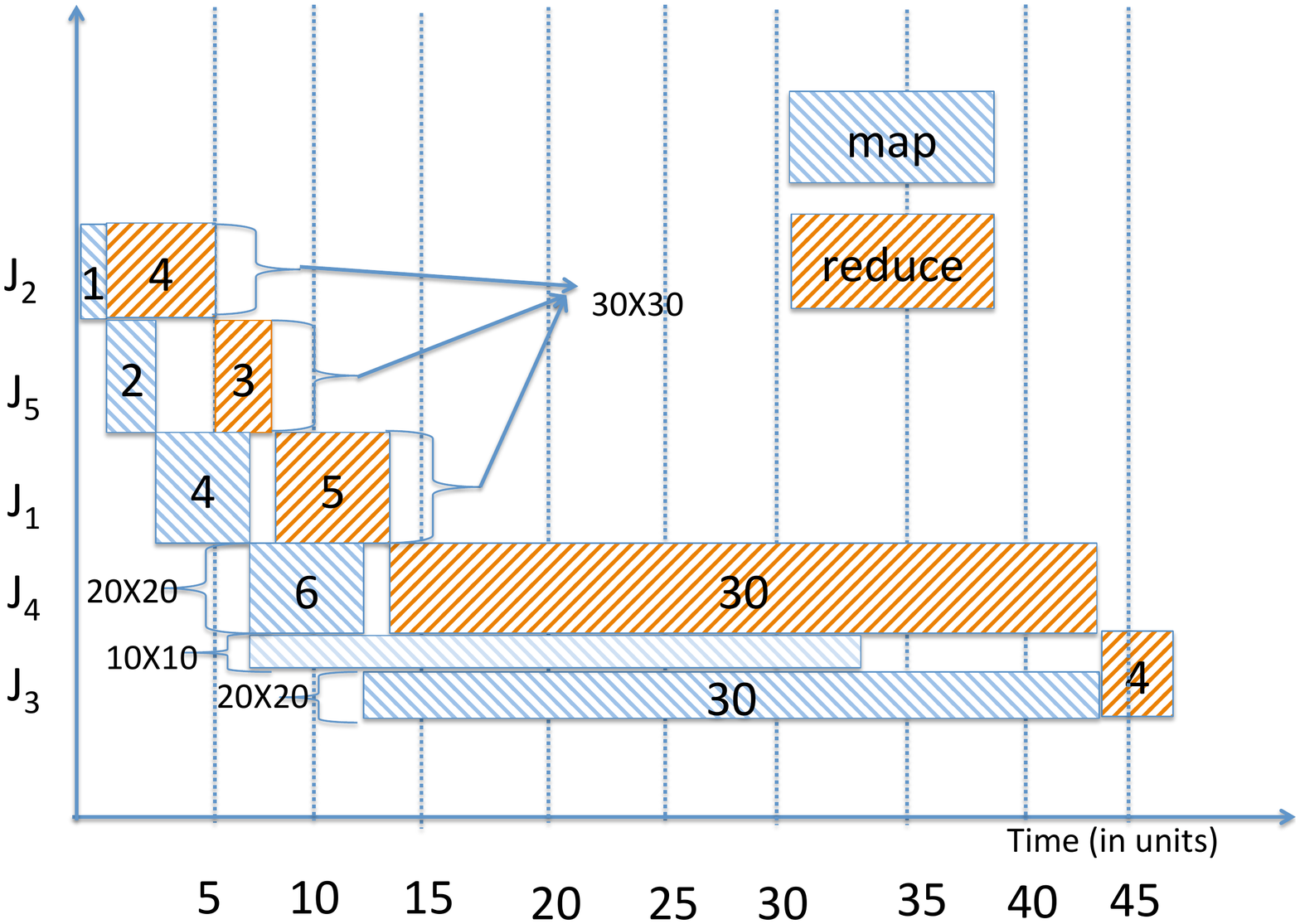}}
\caption{Five MapReduce Jobs Execution in One Cluster by MK$\_$JR}
\end{center}
\end{figure}

\begin{figure} [htp!]
\begin{center}
{\includegraphics [width=0.42\textwidth,angle=-0] {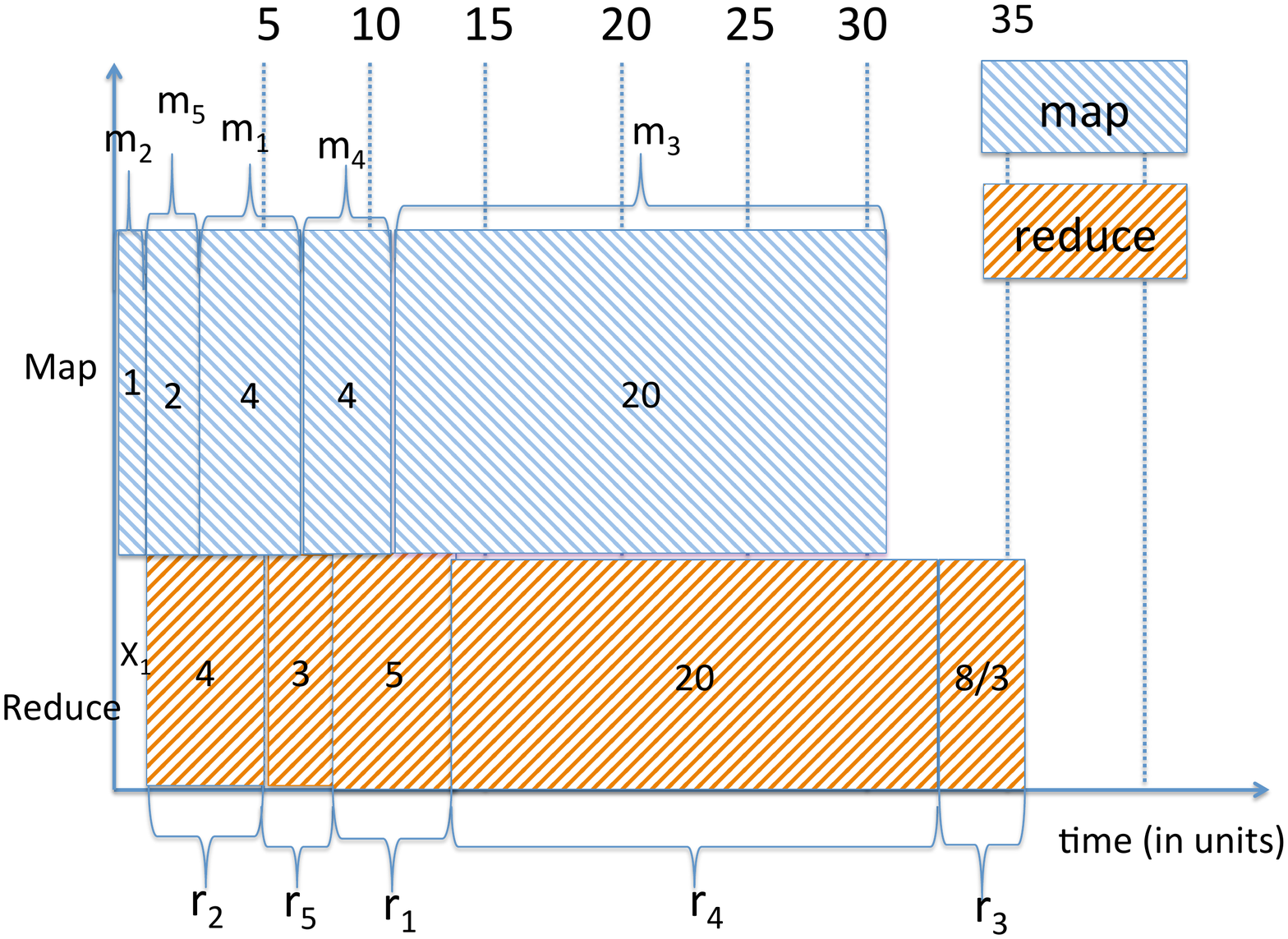}}
\caption{New Result of Five MapReduce Jobs Execution by UAAS}
\end{center}
\end{figure}

\section{Three Algorithms Compared}
In this section, we compare UAAS algorithm with two best known algorithms (BalancedPools and MK$\_$JR) regarding to minimize the makespan of a set of offline MapReduce Jobs. \\
BalancedPools Algorithm [1]: is way to minimize the makespan for offline scheduling proposed in [1], it partitions the Hadoop cluster into two balanced pools and then allocated each job to a suitable pool to minimize the makespan. \\
\textbf {Observation 2. BalancedPool Algorithm does not meet the requirement of Johnson model but just order the job by Johnson rule and is a heuristic algorithm with computational complexity of $O(n^2logn logP)$ where $n$ is the number of jobs and $P$ is the number of MapReduce slots.} \\

MK$\_$JR algorithm [17]: Divide the jobs set $J$ into two disjoint sub-sets $J_A$ and $J_B$. Set $J_A$ consists of those jobs $J_i$ for which $T_i^M<T_i^R$. Set $J_B$ contains the remaining jobs. Sequence jobs in $J_A$ in non-decreasing order of $T_i^M$ and those in $J_B$ in non-increasing order of $T_i^R$. The job order is obtained by appending the sorted set $J_B$ to the end of sorted set $J_A$. \\
\textbf {Observation 3. MK$\_$JR algorithm does not meet the requirement of Johnson model but just order the job by Johnson rule after estimating the map and reduce durations of each job.} \\

The reason that BalancedPools and MK$\_$JR algorithms do not meet the requirement of Johnson model lies that they do not utilize all available MapReduce slots for each job in general case, though they estimate the job ordering by Johnson rule. Therefore, unlike UAAS algorithm, BalancedPools and MK$\_$JR algorithms are suboptimal. \\

\begin{table}
\caption{The example of 5 jobs}
\begin{center}
\begin{tabular}{|l|l|l|l|l|l|}
\hline Job ID &$S_r^M$ &$S_r^R$ & $T_i^M$ &$T_i^R$
\\\hline
\hline $J_1$ & 30&30 & 4 & 5 \\
\hline $J_2$ & 30&30&1 & 4 \\
\hline $J_3$& 20&20 &30 &4\\
\hline $J_4$ & 20&20&6 &30 \\
\hline $J_5$ & 30&30& 2&3\\
\hline
\end{tabular} \\
\end{center}
\end{table}
Table 1 shows an example from [2], where $S_r^M$ and $S_r^R$ is the requested number of slots for map and reduce stage respectively for job $J_i$.
\textbf {Example 1.}  Consider a scenario shown in Table 1 from [1], where the cluster has a configuration of 30$\times$30 MapReduce slots. There are five jobs, among them, $J_1$,$J_2$ and $J_5$ require 30$\times$30 MapReduce slots while $J_3$ and $J_4$ require 20$\times$20 MapReduce slots. The total makespan by MK$\_$JR algorithm is 47 units, visualized in Fig.1. However, if we allow that any job can use all available MapReduce slots in the system when execution (this can be implemented easily in Hadoop, for example by splitting the input files based on available number MapReduce slots), the result is very different from both MK$\_$JR and BalancedPools algorithms. For the same example, in UAAS algorithm job $J_3$ and $J_4$ can use all available $30\times30$ MapReduce slots, then $J_3$ will have Map and Reduce durations (20, $\frac{8}{3}$), $J_4$ will have Map and Reduce durations (4, 20) respectively. Therefore the total makespan will be 35$\frac{2}{3}$ as shown in Fig. 2, where $X_1$=1. This result is smaller (about 31.76$\%$) than the result (47 units) obtained by MK$\_$JR in [1]. The makespan of Pool1 and Pool2 is 39 and 40 time units respectively by applying BalancedPools algorithm, where Pool1 has configuration of 10$\times$10 MapReduce slots and Pool2 has configuration of 20$\times$20 MapReduce slots, and $J_1$,$J_2$ and $J_5$ (short jobs) are with Pool1 while $J_2$ and $J_3$ (longer jobs) are with Pool2. Therefore, the UAAS result is about 12.14$\%$ smaller than the result (40 time units) obtained by BalancedPools algorithm.\\


\textbf{Theorem 2. MK$\_$JR is an 3-approximation algorithm for the makespan optimization in general case.}\\
\textbf{Proof:} Applying the intermediate results from [17] (Equ. (8) in supplementary material for proof of THOREM 1 in [17]),
we have
\begin{equation}
C_{max} \leq (\hat{C_{max}}+ max_{k=1}^{n} {\sum_{i=1}^k {\hat{t_i^M}}}+ max_{k=1}^{n}{\sum_{i=1}^k {\hat{t_i^R}}})
\end{equation}
where $\hat{t_i^M}$ and $\hat{t_i^R}$ is the estimated map and reduce duration for job $J_i$, respectively. Let us define $\sigma$=$\frac{ \max_{k=1}^{n} {\sum_{i=1}^k {\hat{t_i^M}}}+ \max_{k=1}^{n}{\sum_{i=1}^k {\hat{t_i^R}}}} { \hat{C_{max}}}$, the same as in [17],
where $\hat{C_{max}}$ is the theoretical optimal makespan given by Equ. (1)-(2). Considering the worst case that there are two jobs $J_1$ and $J_2$, $T_1^M$=1,~$T_2^M$=$C_0$, and $T_1^R$=$C_0$,~$T_2^R$=$1$; In this case, the optimal order is $J_1$-$J_2 $ and $ \max_{k=1}^{n}{\sum_{i=1}^k {\hat{t_i^M}} }$=$ \max_{k=1}^{n}{\sum_{i=1}^k {\hat{t_i^R}}}$=$C_0$. And $\hat{C_{max}}$=$C_{max}^{opt}$, $C_{max}^{opt} $= $ \max_{k=1}^{n}{\sum_{i=1}^k {\hat{t_i^M}}}$=$C_0$+2 by UAAS algorithm, we have $\sigma$=$\frac{2C_0+1}{C_0+2} \leq 2$. Therefore the approximation ratio of MK$\_$JR is
\begin{equation}
\frac{C_{max}(MK\_JR)}{C_{max}^{opt}}=\frac{C_0+2+\sigma}{C_0+2} =\frac{C_0+2+\frac{2C_0+1}{C_0+2}}{C_0+2} \thickapprox 3
\end{equation}
$\blacksquare$ \\
It worths notice that the worst case is applied for approximation ratio. (1+$\sigma$)-appromixation algorithm where $\sigma$ $\in [0,2]$, should be called 3-approximation algorithm since $\sigma$ is 3 in the worst case.
\\
Based on previous results, we have the following observation. \\
\textbf{Observation 4. BalancedPools and MK$\_$JR algorithms are suboptimal regarding to minimizing the makespan, they may not have the minimum makespan for a set of jobs; applying Theorem 1 to single Hadoop cluster always has optimal total makespan for a set of jobs.}\\

\textbf{Theorem 3. Given a homogeneous environment where the Hadoop configurations of slave nodes are identical, the job order $\phi_1$ produced by UAAS for a batch of jobs are independent of the number of slave nodes ($N$) but depends on the total number of available MapReduce slots ($|S^M|$, $|S^R|$), and is stable with regarding to the change of the total number of slave nodes.} \\

\textbf{Proof:} Let us set the execution durations of map and reduce stages for a given job $J_i$ under a given configuration of Hadoop cluster with $|S^M| \times |S^R|$ MapReduce slots, as $T_i^M$ and $T_i^R$, respectively. If the MapReduce slots configuration of Hadoop cluster is changed to $|S_x^M| \times |S_x^R|$ and set $\rho_0$=$\frac{|S^M|}{|S_{x}^M|}$. Applying UAAS algorithm, the execution duration of map and reduce stage for a given job $J_i$ will change to $T_i^{M'}$ and $T_i^{R'}$. And we have
\begin{equation}
T_i^{M'}=T_i^M \frac{|S^M|}{ |S_{x}^M|}=T_i^M \rho_0
\end{equation}
\begin{equation}
T_i^{R'}=T_i^R \frac{|S^R|}{ |S_{x}^R|}=T_i^R \rho_0
\end{equation}

This means execution duration of map and reduce stage for each job will change proportional to $\rho_0$ but their relative relationship (ordering by their durations) will not change. Therefore the job order of UAAS is stable with regarding to the change of the total number of slave nodes. $\blacksquare$ \\

\textbf{Observation 5. The the job ordering of MK$\_$JR and BalancedPools is not stable when the total number of slave nodes changes.}

Let us consider the example given in [17]. There is a Hadoop cluster with 5 nodes, each configured with 2 map and 2 reduce slots. Let $J_1$ be defined as follows: Map stage duration is 9 and requires 10 map slots. Reduce stage duration is 10 and requires 1 reduce slot. Let $J_2$ be defined as follows: Map stage duration is 11 and requires 8 map slots and reduce stage duration is 15 and requires 1 reduce slot. In this case, the optimal job scheduling order by UAAS is $J_2$-$J_1$, and their corresponding map and reduce duration is (8.8,1.5) and (9,1) respectively by utilizing all 10 MapReduce slots in each stage, with the makespan of 18.8. The job order produced by MK$\_$JR is $J_1$-$J_2$ with the makespan of 35, which is about 86.17$\%$ larger than optimal result.
Now, if one node fails, then there are only 4 nodes left with
8 map and 8 reduce tasks available in the cluster. In this case,
the optimal
job scheduling by UAAS is still $J_2$-$J_1$, however, their corresponding map and reduce duration is (11.25,1.25) and (11, 1.875) respectively by utilizing all 8 MapReduce slots in each stage, with makespan of 23.5. The job order generated by MK$\_$JR keeps the same,
i.e., $J_1$-$J_2$, with makespan of 43, about 82.97$\%$ larger than the optimal. \\
Notice that BalancedPools algorithm has following results. When there are 5 nodes, $J_1$ with duration (9,10) will be put into Pool1 with 2 nodes of 4 MapReduce slots and $J_2$ will be allocated to Pool2 with 3 nodes of 6 MapReduce slots. Then $J_1$ will have duration (22.5, 10) and $J_2$ will have duration (14.67,15). If one node fail, $J_1$ still with Pool1 and $J_2$ with Pool2; $J_1$ and $J_2$ will have duration (32.5, 10) and (37, 10) respectively. In either case, BalancedPools is far from optimal results. \\

\textbf{Theorem 4. Let $\rho$ be the ratio of map slots to reduce slots, i.e., $\rho$=$\frac{|S^M|}{|S^R|}$. The optimal configuration of $\rho$ for makespan $C_{max}$ depends on the total number of slots ($|S^M|$, $|S^R|$), MapReduce workload as well as its job submission order $\phi$.}\\
\textbf{Proof:}

\begin{eqnarray}
C_{max} &=& \max_{k=1}^{n} {\sum_{i=1}^k {{T_i^M}}}+ \max_{i=k}^{n}{\sum_{i=k}^n {{T_i^R}}} ~ ~from [17] \nonumber \\
&=& \max_{k=1}^{n} ({\frac{1}{|S^M|} \sum_{i=1}^k \sum_{j=1}^{|J_i^M|} {{t_{i,j}^M}}}+ {\frac{1}{|S^R|} \sum_{i=k}^n \sum_{j=1}^{|J_i^R|} {{t_{i,j}^R}}})\nonumber \\
&=& \frac{1}{|S^R|} \max_{k=1}^{n} ({\frac{1}{\rho} \sum_{i=1}^k \sum_{j=1}^{|J_i^M|} {{t_{i,j}^M}}}+ {\sum_{i=k}^n \sum_{j=1}^{|J_i^R|} {{t_{i,j}^R}}})
\end{eqnarray}
This means the optimal configuration of $\rho$ for makespan $C_{max}$ depends on the total number of slots ($|S^M|$,$|S^R|$) MapReduce workload ($t_{i,j}^M$, $t_{i,j}^R$ ) as well as its job submission order $\phi$(=$\{1$,..$n$\}).
$\blacksquare$ \\
When the workload and job order are fixed, it is obvious that larger number of total number of MapReduce slots will lead to smaller value of $C_{max}$. This is consistent with Theorem 1 and UAAS algorithm to utilize all available MapReduce slots ($|S^M|$, $|S^R|$).
\section{Conclusion}
Observing that there are quite a few algorithms on minimizing makespan of multiple MapReduce jobs and these algorithms are heuristic or suboptimal. In this paper, we proposed an optimal approach called UAAS algorithm to minimize the makespan of a set of MapReduce jobs. The proposed algorithm meets the requirements of classical Johnson algorithm and therefore is optimal with regarding to the makespan. We also conducted extensive tests in real Hadoop environment to validate our theoretical results by benchmarks provided in [13][14]. Because this is a short paper, we do not provide the test results yet. There are future research directions such as considering minimizing the makespan of online MapReduce jobs and minimizing the total completion time and total flow time of a set of Mapreduce jobs.

\end{document}